\documentclass[a4paper,12pt]{article}
\usepackage{epsfig,amssymb,amsmath}
\usepackage{graphicx}
\usepackage{amsfonts}

\begin{document}

\newcommand{\be}{\begin{equation}}
\newcommand{\ee}{\end{equation}}
\newcommand{\bea}{\begin{eqnarray}}
\newcommand{\eea}{\end{eqnarray}}
\newcommand{\bean}{\begin{eqnarray*}}
\newcommand{\eean}{\end{eqnarray*}}
\newcommand{\nn}{\nonumber}
\font\upright=cmu10 scaled\magstep1
\font\sans=cmss12
\newcommand{\ssf}{\sans}
\newcommand{\stroke}{\vrule height8pt width0.4pt depth-0.1pt}
\newcommand{\C}{\mathbb{C}}
\newcommand{\CP}{\mathbb{CP}}
\newcommand{\Z}{\mathbb{Z}}
\newcommand{\half}{\frac{1}{2}}
\newcommand{\quart}{\frac{1}{4}}
\newcommand{\bphi}{\bar{\phi}}
\newcommand{\bPhi}{\bar{\Phi}}
\newcommand{\bz}{\bar{z}}
\newcommand{\bZ}{\bar{Z}}
\newcommand{\pr}{\partial}
\newcommand{\bm}{\boldmath}
\newcommand{\I}{{\cal I}} 
\newcommand{\Lag}{{\mathcal L}}
\newcommand{\M}{{\cal M}}
\newcommand{\N}{{\cal N}}
\newcommand{\V}{{\cal V}}
\newcommand{\e}{\varepsilon}
\newcommand{\g}{\gamma}
\newcommand{\Tr}{{\rm Tr}}

\thispagestyle{empty}
\vskip 3em
\begin{center}
{{\bf \LARGE Approach to nuclear cross sections via
classical Skyrmion scattering}}\\[10pt] 

{{\bf \large A proposal}}
\\[15mm]

{\bf \large N.~S. Manton\footnote{email: N.S.Manton@damtp.cam.ac.uk}} \\[20pt]

\vskip 1em
{\it 
Department of Applied Mathematics and Theoretical Physics,\\
University of Cambridge, \\
Wilberforce Road, Cambridge CB3 0WA, U.K.}
\vspace{12mm}

\abstract
{By analogy with heavy ion collisions, which can be modelled by a
classical hydrodynamics, it is proposed that the differential
cross section for collisions of smaller nuclei can be
calculated from the classical, numerical scattering data for
Skyrmions. It is also suggested that the numerical data for the
outgoing particles should be classified into bins, as is done in
nuclear physics experiments. 
}

\end{center}

\vskip 150pt
\vskip 1em

\vfill
\newpage
\setcounter{page}{1}
\renewcommand{\thefootnote}{\arabic{footnote}}

\section{Introduction} 

Thought of in terms of nucleons and quarks, heavy ion collisions are
highly quantum mechanical. But for a long time, they have been modelled
by an essentially classical hydrodynamics of strongly interacting
matter. Because the collision is so much quicker than the time for a
nucleus to rotate, it can be assumed that the nuclear orientation
has a definite value in each collision \cite{HeavyIon}. For
unpolarised nuclei, even if they have a non-zero spin, the orientations have
a uniform distribution on SO(3). When the collision is modelled
theoretically, all orientations need to be considered, and the results
averaged. Similarly the impact parameter $b$ can be assumed to have a
definite value in each collision, as a step of $\hbar$ in the orbital
angular momentum of the heavy ion pair corresponds to a classical change
of impact parameter that is much less than the nuclear radius. In a
theoretical model, the impact parameter runs from zero to a cut-off
a little larger than the nuclear diameter, with a uniform distribution
in the beam transverse area, i.e. a distribution increasing linearly with $b$.

More precisely, classical orientational behaviour is justified provided
the collision time is much shorter than the rotation time for
unit spin, which is $2\pi I/\hbar$ if the nuclear moment of inertia is
$I$ \cite{HeavyIon}. Similarly, the criterion justifying that the
collision has a classical impact parameter is that the orbital angular
momentum $Mvb$ is much larger than $\hbar$, where $M$ is the nuclear
mass and $v$ the collision velocity of each nucleus. For $b$ to be a
small fraction of the nuclear radius, the velocity $v$ needs
to exceed some small critical value, but this is easily satisfied
in relativistic heavy ion collisions.

These considerations raise the possibility that Skyrmion collisions at
substantially lower energies, modelling collisions of quite small
nuclei, even a pair of nucleons, could be treated in a similar, classical
way. For the author's previous discussion of this, see ref.\cite{Man}.
Classical Skyrme field dynamics would play the role of the heavy ion
hydrodynamics. The Skyrme field is the nonlinear isospin
triplet of pion fields possibly supplemented by an $\omega$-meson
field \cite{Skbook}, giving more dynamical detail than just the baryon
density together with its velocity and perhaps temperature.
Braaten had earlier partially implemented this approach for
low-energy collisions, though computer speeds at the time were
not sufficient to do this fully \cite{Bra}.

One needs to check if the classical dynamical results are justified. The
most demanding case is the elastic scattering of two $B=1$ Skyrmions (where
$B$ denotes baryon number). Let us work in Skyrme units, where the
length unit is essentially 1 fm (and the speed of light is 1), and the
energy unit is about 6 MeV. Crucially, $\hbar$ is approximately 30. This is
determined by the spin splitting of states of some of the larger Skyrmions
calibrated to nuclear states, for instance of Carbon-12, which in turn
depends on the classical Skyrmion moments of inertia \cite{Skbook}. With
D. Foster, I found that a typical angular velocity of a $B=1$ Skyrmion in a
spin-half nucleon state is about 0.3, so the rotation time is 20 \cite{FM}.
Suppose that in the CM frame the colliding Skyrmions each have
speed $v$. The collision time is $1/2v$, so one requires $1/2v \ll 20$
for the orientations to be treated as having definite values. Thus
$v \gg 0.025$. This is easily achieved in non-relativistic
collisions with $v$ in the range $0.2 - 0.5$. Larger velocities
would be undesirable, because they would be well above the
pion production threshold.

Larger-$B$ Skyrmions have larger moments of inertia, and the classical
criterion is easier to satisfy, because the rotation angular velocity
is smaller but the collision time little changed.

Skyrmions also have an isospin orientation. For $B=1$, spin and
isospin angular velocities are the same. For larger Skyrmions, modelling stable
nuclei with unequal numbers of protons and neutrons, the isospin grows
quite slowly with $B$, but the isospin moment of inertia is
roughly proportional to $B$, so the angular velocity is
decreasing. Isospin orientation can therefore also be treated as
having a definite value in a single collision.

What about impact parameter for $B=1$ Skyrmion collisions? Let $b$ be
the impact parameter corresponding to one unit of orbital angular
momentum, $\hbar$, and suppose the desired impact parameter resolution
is $b = 0.2$. Then $Mvb = \hbar$. Now, in Skyrme units,
$M \simeq 1.2 \times 12\pi^2 \simeq 140$. Thus $v \simeq 1$. In other
words, the classical spatial resolution is rather poor unless the
velocity is large — at least 0.5. Fortunately, this is the worst
case. If one or both of the Skyrmions have higher $B$, the velocity
can be smaller and the classical impact parameter makes more precise
sense.

\section{Differential cross section calculations}

Let us now consider how to practically calculate the differential
cross section in Skyrmion scattering. Nuclear data is invariably in
terms of a differential cross section, so this is what is needed to
compare Skyrmion and nuclear scattering. There have been a number
of numerical simulations of Skyrmion scattering \cite{VWWW, AKSS,
BS, GP, AHP, FK}, especially for two $B=1$ Skyrmions. Interesting
scattering has been established, including $90^\circ$ scattering in
the attractive channel. However, so far the simulations have not
been done for enough initial orientations and impact parameters to calculate
a differential cross section. One of the closest is the calculation of
Amado et al. \cite{AHP}, where the scattering angle $\theta$ as a function
of impact parameter $b$ was found for a limited selection of initial
orientations. In each channel, the cross section can be obtained from the
inverse of the derivative of $\theta$ w.r.t $b$. Essentially this reduces the
problem to planar scattering, similar to Samols' calculation for vortices
\cite{Sam}, and Leese's calculation for $\sigma$-model lumps in a
particular channel originally noted by Ward \cite{Lee}. Interestingly,
Amado et al. worked with an $\omega$-meson variant of the Skyrme model
that makes numerical Skyrmion scattering more stable than
in Skyrme's original model with its term quartic in derivatives.

Our ideas in the past have depended on creating a 2-Skyrmion
moduli space, with a metric and potential \cite{AtMan}, and then calculating a
quantum differential cross section by precise solution of the Schr\"odinger
equation on moduli space for incoming plane waves, or by a partial
wave analysis. So far there's been little progress with this, except by
making some substantial truncations of the degrees of freedom. Quantum
scattering calculations for vortices and BPS monopoles are slightly
easier, and progress has been made by Samols in the vortex case, and
by Schroers for monopoles \cite{Sch}.

I propose that one can go further with a different approach to the
cross section, inspired by what experimenters do. The suggestion is to
do random classical scatterings, rather than attempting to establish
classical scattering angles as smooth functions of initial data. The
orientations should be chosen randomly from a uniform
distribution, and impact parameters chosen randomly, uniformly covering a
circular area of impact. Amado et al. put the impact parameter cut-off
at 2.8 fm (about 2.8 Skyrme units), because the scattering is
negligible beyond this. More interesting is how to classify the
outgoing Skyrmions. The experiments have detectors of finite size, so
they detect outgoing particle angles in bins of a few degrees angular
width. I suggest binning the numerical scattering output similarly.
The near-forward scattering data will be unreliable, because of the
impact parameter cut-off, but otherwise this approach should work, and
ease the comparison with experimental results. An approach like this
to classical BPS monopole scattering has been implemented by
Temple-Raston and Alexander \cite{TA}.

For randomly oriented (unpolarised) $B=1$ collisions, the averaged
result only depends on the outgoing polar angle (measured away from
forward scattering), but for more general collisions, perhaps with
polarised incoming particles, one should work with bins of solid
angle, as in the experiments. Random orientations imply that only the
magnitude of the impact parameter needs to be varied. Also,
because of isospin symmetry, the orientation of one Skyrmion can be
fixed, while the other is varied.

For elastic 2-particle collisions, there is the simple formula for the
estimated differential cross section,
\be
\frac{d^2\sigma}{d\Omega} = \frac{N(\Omega)}{F \times \Delta\Omega} \,,
\label{1+1cross}
\ee
where the right-hand side is the number of particles $N(\Omega)$ reaching
the bin at solid angle $\Omega$, divided by the product of the flux
$F$ (the total number of incoming particles per unit area) and the
solid angle size $\Delta\Omega$ of the bin.

There is a generalisation of this estimated differential cross section in
the interesting but more difficult case of the break-up differential
cross section in proton-deuteron collisions. Recent experiments on
this reaction are discussed in refs.\cite{Koz1,Koz2}. Here there are
three outgoing particles (two protons and a neutron), and the two protons
are detected. The initial collision energy is fixed in a particular
experiment. The outgoing solid angles of the detected protons are
classified into bins, and the differential cross section depends on the
proton count in each pair of bins but also on a further
kinematic variable $S$, essentially the ratio of the energies of the
outgoing protons. Knowing the solid angles and $S$, one can
reconstruct the momenta of both protons and the neutron.
The variable $S$ and its binning does not have a simple
formula, but is explained graphically in ref.\cite{Koz1}.

Theoretically, the differential cross section has two infinitesimal
solid angles and $dS$ in the denominator. But experimentally, this is
replaced by the count of protons into the bins at solid angles
$\Omega_1$ and $\Omega_2$, and the bin at $S$. Importantly, the
proton detectors can measure the proton energy and hence infer
$S$. For more details, see refs.\cite{Koz1,Koz2}. The estimated
differential cross section is
\be
\frac{d^5 \sigma}{d\Omega_1 d\Omega_2 dS}
= \frac{N(\Omega_1,\Omega_2,S)}{F \times \Delta\Omega_1 \Delta\Omega_2
\Delta S} \,,
\label{1+2cross}
\ee
where $F$ is again the total number of incoming particles per unit
area, $\Delta\Omega_1$ and $\Delta\Omega_2$ are the solid angle
bin sizes, and $\Delta S$ a chosen bin size for $S$.
This is an idealised version of a formula in ref.\cite{Koz1}, with the
measurement efficiency set to unity. This result is then compared
with theoretical calculations of the same differential cross section,
and good agreement is found. Note that the break-up cross section is
just part of the total -- additionally there is elastic
proton-deuteron scattering.

The same reaction could be studied by numerically
colliding $B=1$ and $B=2$ Skyrmions. The total energy can be fixed, but
needs to be sufficient for the $B=2$ Skyrmion to routinely break up.
I suggest that the Skyrmion output data is treated similarly as in
the experiments, by classifying it into bins of solid angle and $S$.
Multiple scattering runs, with varying impact parameter and Skyrmion
orientations, are required. The impact parameter should have some cut-off, and
all incoming orientations treated equally (for unpolarised protons
and deuterons). It may be permissible to bias the density of
orientations away from uniform to explore regions of particularly
small cross section, or regions of particular interest, provided the
normalisation of the count is adjusted appropriately.

\section{Symmetric Space Star} 

The region of special interest for proton-deuteron break-up is the
symmetric space star, where (in the CM frame) the outgoing particles
have equal speeds and form an equilateral triangle orthogonal to the
initial collision direction \cite{Wil,Wit}. The differential cross section
around the space star is not accurately fitted by any theory of
nuclear forces at present, and it has been said that a quite new idea is
needed to explain this anomaly if it is real. But the experiments
finding this anomaly are quite old and at rather low energy, with
beams of 10 — 20 MeV. The anomaly is less pronounced at 65 MeV \cite{Wit}.

Skyrmions may provide such a new idea, because unlike point nucleons,
they allow the interchange of energy and matter in close collisions
\cite{FK}. Also, we know special initial conditions for Skyrmions that
lead to the space star. These can be derived within the rational map ansatz
for Skyrmions \cite{HMS} using rational maps with $C_3$ (but not $D_3$)
symmetry. Then, to get the differential cross section one must vary
the initial impact parameter and orientations, and set up bins around
the space star output. The experiments suggest it is important to look at the
differential cross section much more widely than this, to calibrate properly,
and compare with known theoretical, nuclear-physics calculations.

It is interesting to know if the space-star anomaly actually persists
in higher-energy collisions. The most recent experiments \cite{Koz1,Koz2}
have apparently not yet explored the symmetric space star -- at least, it
is not explicitly mentioned. They have either a proton beam on a deuteron
target at rest, or the reverse of this (inverse kinematics). It's easy
to shift to the CM frame, simply by a velocity shift in non-relativistic
kinematics (selected results in relativistic kinematics are given in
ref.\cite{Wil}). The space star occurs in the lab frame at a forward polar
angle of $\arctan{\sqrt{2}} \simeq 55^\circ$ for normal, non-relativistic
kinematics, or $\arctan({1/\sqrt{2}}) \simeq 27^\circ$ for
inverse kinematics. These angles are independent of beam velocity,
provided one ignores the deuteron break-up energy of 2 MeV compared
with the impact energy, which is of order 100 MeV in the recent experiments.
Ref.\cite{Koz2} mentions that the currently explored range of polar
angles for normal kinematics is $13^\circ$ — $33^\circ$.

\section{Planar scattering}

For completeness, we recall some work on classical soliton
scattering in two space dimensions. For solitons scattering
elastically in a plane, there is just one scattering angle and this
is a function of impact parameter. Here, some differential cross
sections have been calculated.

For the planar scattering of vortices \cite{Sam} and lumps \cite{Lee},
and the effective planar scattering of Skyrmions \cite{AHP} in the
attractive channel, the graph of scattering angle against
impact parameter is explicitly or implicitly known, though not necessarily
analytically. From this graph the differential cross section can be
derived from the slope. For the BPS vortices and lumps, the result
has been obtained by calculating geodesics on
the 2-soliton moduli space. The $90^\circ$ scattering at zero impact
parameter is the planar analogue of symmetric space star formation,
so the cross section at this angle is particularly interesting. In
the case of lumps it depends on the asymptotic behaviour of an
analytic description \cite{Lee}, but this has so far resisted
calculation. It would be interesting to understand this differential cross
section at $90^\circ$ in terms of the moduli space curvature. The
differential cross section is related to geodesic deviation, so such
an understanding should be possible.


\section*{Acknowledgements}

This note was stimulated by a query from Shengquan Tuo concerning
the detection of alpha-cluster shapes of Oxygen-16 and Neon-20 in
heavy-ion collisions. I thank Paul Sutcliffe for drawing
attention to potential advantages of the $\omega$-meson variant
of the Skyrme model. This work has been partially supported by STFC
consolidated grant ST/P000681/1.


\end{document}